\documentstyle[preprint,revtex]{aps}
\begin{document}
\draft

\def\ftnote#1{\nolinebreak\renewcommand{\baselinestretch}{1}%
\nolinebreak\footnote{#1}
\renewcommand{\baselinestretch}{1.5}\nolinebreak}

\def\gnufig#1{\begin{figure}[#1]\begin{center}\vspace*{-8mm}}
\def\gnucaption#1#2#3{\end{center}\vspace*{-6mm}\caption[#2]{#3}\label{#1}\end{figure}}
\newcommand{\grad}{\nabla}
\newcommand{\del}{\partial}
\newcommand{\up}{\uparrow}
\newcommand{\dn}{\downarrow}
\newcommand{\be}{\begin{equation}}
\newcommand{\ee}{\end{equation}}
\def\refp#1{(\ref{#1})}
\def\bm#1{\mbox{\boldmath $#1$}}
\def\avg#1{\mbox{$\langle #1 \rangle$}}
\def\etal{{\em et~al.\ }}

\begin{title}
New Dynamic Monte Carlo Renormalization Group Method
\end{title}

\author{Martin-D. Lacasse$^{1}$, Jorge Vi\~nals$^{2}$ and Martin Grant$^{1}$}

\begin{instit}
$^{1}$ Centre for the Physics of Materials and\\
Department of Physics, McGill University,\\
Rutherford Building, 3600 University Street,\\
Montr\'eal, Qu\'ebec, Canada H3A 2T8,\\
$^{2}$ Supercomputer Computations Research Institute, B-186,\\
Florida State University, Tallahassee, Florida 32306-4052.\\
\end{instit}

\receipt{June 12, 1992}

\begin{abstract}

The dynamical critical exponent of the two-dimensional spin-flip
Ising model is evaluated by a Monte Carlo renormalization
group method involving a transformation in time. The results
agree very well with a finite-size scaling analysis performed
on the same data. The value of $z = 2.13 \pm 0.01$
is obtained, which is consistent with most recent estimates.

\end{abstract}


\narrowtext

\section{Introduction}
The dynamics of the two-dimensional spin-flip Ising model
remains a source of interest and controversy.
Near the critical point, not only does the correlation length
$\xi$ diverge following $\xi \sim |T - T_c|^{-\nu}$, where $T_c$ is the
critical temperature and $\nu$ a critical exponent, the correlation time
$\tau$ also diverges due to critical slowing down.
This divergence can be characterized by the dynamic critical exponent
$z$, where $\tau \sim \xi^z$, or equivalently, \mbox{$\tau \sim |T -
T_c|^{-\Delta}$}, where  $\Delta = \nu z$.
The dynamic scaling hypothesis asserts that,
in the long time limit, all times scale with
this diverging time scale, so that a dynamical universality class is
characterized by, amongst other things, this critical exponent $z$.
Since the particular value of $z$ is common to all members of a
universality class, it is an important and fundamental
quantity.  Here we use a novel Monte Carlo renormalization group method
to estimate $z = 2.13 \pm 0.01$ for the two-dimensional spin-flip Ising
model, which is in the universality class of model A \cite{Halperin77},
where a nonconserved scalar order parameter is the only dynamical mode.

Many authors
\cite{Stoll73,Achiam78,Chui79,%
Tobochnik81,Takano82,Katz82,Williams85,Miyashita85,%
Tang87,Ito87,deAlcan87,Jan88,Landau88,%
Racz75b,Angles82,Bolton82,%
Miyashita83,Kalle84,Mori88,Poole90,MacIsaac92,%
Yahata69,Racz76,Bausch81,Mazenko81,Rogiers90,Wang90}
have attempted to evaluate $z$
by a variety of techniques,
giving values that are not always consistent with each other, even when
considering the given errors.
There are many possible explanations of these discrepancies,
and we shall consider some of them.
First, however, we shall discuss some lower bounds.
For model A, mean field theory  predicts $z = 2 - \eta$,
where $\eta$ is the critical exponent describing the power-law decay
of the correlation function.
Using a generalized Langevin-equation approach, Schneider \cite{Schneider74}
showed this was a lower bound on $z$.
Of course, the mean-field result is correct at and above the upper
critical dimension $d_u = 4$, below which the $\epsilon= 4-d$ expansion
\cite{Halperin77}
gives $z = 2 + 0.01345 (4-d)^2 - 0.002268 (4-d)^3 + {\cal O}(4-d)^4$.
The lower critical dimension of model A is $d_\ell = 1$, and
Bausch \etal \cite{Bausch81}
have calculated the $d=1+\epsilon$ expansion of the kinetic drumhead
model to be:
$z = 2 + (d-1) - \frac{1}{2}(d-1) ^{2} + {\cal O}(d-1)^3$.
More recently, one of us \cite{Grant89}
argued that the dynamic critical exponent should be larger than the reciprocal
of the exponent for domain growth, yielding $z \geq 2$.

Of the different techniques used for the determination of $z$,
no method seems to have proved better than others.
Moreover, a given method does not always
yield consistent results. For example, high-temperature expansions
of different orders
apparently converge to different values whether the expansion
is computed for spin-spin time correlations
\cite{Yahata69,Racz76}
or magnetization correlations \cite{Rogiers90,Wang90}.
On the other hand, Monte Carlo techniques have systematic
errors which can be quite difficult to evaluate. For example, it is generally
believed \cite{Williams85,Kikuchi86,Angles82,Ceccatto86} that
the dynamic universality class of the
spin-flip Ising model is insensitive to the algorithm used
as long as the updating algorithm is local \cite{Swendsen87,Wolff89}.
Some results obtained in one dimension \cite{Williams85a,Cordery81}
support this hypothesis. However, studies in three dimensions using Creutz's
deterministic microcanonical dynamics \cite{Brower88} yield a value
of $z$ that ``agrees with the Monte Carlo measurements as they agree among
themselves'' but is nevertheless
slightly higher than most recent estimates (e.g. \cite{Brower88,Wansleben91}
and references therein).
This discrepancy could be due to the methods of estimation
of $z$ rather than to the algorithm,
but to our knowledge, there
exists no systematic analysis testing
the limits of validity of the dynamic universality class hypothesis.
This is a potentially  important issue, since nonlocal acceleration
algorithms are being developed \cite{Swendsen87,Wolff89}
which have exceedingly small values of $z$.

Besides the algorithm, another source of systematic
error is from critical slowing down itself.
While such effects are well known and are typically incorporated
\cite{Zwanzig69,Ferrenberg91} to consistently estimate $z$,
the existence of long
time correlations can potentially induce subtle
couplings of correlations in pseudo-random number generators.

It has also been noted
\cite{Racz76,Racz75} that the linear relaxation critical exponent of
the magnetization $M$ can differ from the nonlinear one.  The nonlinear
relaxation time $\tau$ is defined as $ \tau = \int_0^\infty
(M(t) - M(\infty))/(M(0) - M(\infty)) d\,t $ where the
denominator is for normalization.  If the integrand is
written as a sum of exponentials, $\sum_i w_i e^{-t/\tau_i}$, then
$\tau$ will simply be the weighted sum of the partial relaxation
times.  The limit when $M(0) \rightarrow M_{eq}$ , i.e.\ , as
the system is left to relax from a value near its equilibrium value,
defines the linear relaxation time.  It characterizes a relaxation
process that does not include the scaling of the order parameter
itself.
R\'acz \etal \cite{Racz76,Racz75} used scaling arguments to suggest
that the nonlinear critical exponent $\Delta_A^{(nl)}$ of an observable
$A$ can be related to the linear critical exponent $\Delta_A$ by the
following relation $\Delta_{A}^{(nl)} = \Delta_A - \beta_A $
where $\beta_{A}$ describes the scaling of the
quantity $A$ with respect to temperature.  Mori \etal \cite{Mori88} in
$d=2$, and Chakrabarti \etal \cite{Chakrabarti81} in three dimensions,
reported observations of such a relation.

Current techniques used in Monte Carlo simulations sometimes involve
nonlinear response in terms of the definition given above.  Some
authors \cite{Jan88,deAlcan87,Bolton82,Kalle84} have used the relaxation
of the order parameter in systems prepared at zero temperature
when put in contact with a heat bath
at some temperature near $T_c$. On the other hand, other techniques
\cite{Stoll73,Tobochnik81,Takano82,Katz82,%
Williams85,Miyashita85,Tang87,Ito87,Jan88,Landau88}
involve the measurement of some time correlations in critically
equilibrated systems.  Presumably there could be differences
involving transient dynamics in these methods.
However, there is no real distinction made in
the literature concerning possible discrepancies between these various
methods.  It is not clear yet
how important the distinction between linear and non-linear relaxation
times is, and further investigations should clarify this situation.

Finite-size effects can be important in
simulations of critical systems, e.g., the effect of using the infinite
system critical temperature can induce systematic errors in small
systems.  Although these effects can be exploited by
finite-size scaling, very small systems may not be in the
range where scaling applies. Finite systems also introduce the concept
of {\em ergodic} time \cite{Miyashita83,Binder88}, i.e., the mean
lifetime of the system in one of its broken symmetry states. While infinite
systems at $T_c$ have a vanishing order parameter, finite systems  of
size $L$ have nonzero values $\pm |M^L_c|$ between which the system has
spontaneous transitions. Those transitions introduce a large effect in
time correlations as discussed below.

Finally, it should be noted that experiments have been done on systems
thought to be in the universality class of model A.  The few results
\cite{Racz75a,Hutchings82} of which we are aware yielded values of $z$
well below 2.  However the experiments are difficult,
and the microscopic processes involved in the systems are various, so
that it may be that the experiments do not probe the problem of
interest herein.

\section{Method}

The Monte Carlo renormalization group (MCRG) was introduced by Ma
\cite{Ma76},
and developed by others, especially Swendsen \cite{Swendsen83}.
For critical dynamics, the method was extended by Tobochnik, Sarkar, and
Cordery \cite{Tobochnik81}, and others
\cite{Katz82,Jan88,Kalle84,Williams85}. MCRG allows one to use
the self-similarity in critically
equilibrated systems by analyzing the effect of a controlled change of
length scales on correlation functions.
When length scales are changed by a factor of $b$, by some
suitable blocking of $b^d$ spins to one renormalized spin, the correlation
length changed by $\xi \rightarrow \xi /b$.
This implies that time scales are changed by a factor of $\tau
\rightarrow \tau / b^z$, from which
the dynamic critical exponent can be estimated.
Unfortunately the method is self-consistent in that there is no proof
the system approaches a fixed point under the renormalization group.
Thus it is essential that checks are made that, after several levels of
RG, scaling is consistently observed.
Here, we generalize the usual procedure of blocking in space, by blocking
in time $t$.

Consider the Hamiltonian of an Ising-like system
\begin{equation}
{\cal H} = \sum_\alpha K_\alpha S_\alpha
\end{equation}
where the $K_\alpha$'s are the coupling constants including
temperature, the $\alpha$ index runs over all $i=1,2, \cdots, N$ spins
for nearest-neighbor, next-nearest neighbor (and so on) interactions,
and the $S_\alpha$'s are generalized spins made
of specific products of spins $\sigma_i= \pm 1$ on each site.
For example, the Ising model has
$K_1 = -J/k_BT$, and $K_{\alpha > 1} = 0$ where $J$ is the coupling constant,
$k_B$ is Boltzmann's constant and $T$ is temperature
and $S_1 = \sigma_i \sigma_j$
such that the sum is restricted so that $i$ and $j$ are nearest neighbors.
Here we will consider the two-dimensional Ising model on a square
lattice, and apply a renormalization-group transformation repeatedly to
this evolving system.
As mentioned above, a typical numerical MCRG
transformation \cite{Ma76} is to ``block'' by a length rescaling factor
of $b$.  A block of $b^d$ spins is transformed into a renormalized spin
by majority rule of the spins in the block, with a random outcome on ties.
The resulting renormalized
Hamiltonian is assumed to be expressible in terms
of another short-range Ising-like system, with more
$S_\alpha$ terms contributing.  Numerically, the approximation results
from the fact that the number of spins remaining after $m$ blockings is
$N/b^{md}$, thus coupling constants for interactions are truncated, if
they involve larger length scales than the entire system.  Thus, for
equilibrium properties, a suitably chosen transformation operator will
change the Hamiltonian ${\cal H}(\bm{K})$ by moving the parameter
vector \bm{K} to some other point in $\bm{K}$-space.

For critical dynamics the situation is analogous.  Say one begins with
the Ising model again, with the dynamics of single-spin flips, where
each evolving configuration depends only on the previous one, i.e., a
Markov process. Under the blocking transformation mentioned above, it
is expected that the original master equation which is ``short-range''
in time is changed to a non-Markovian equation which has memory
over some small time scales.

Our original contribution here is to introduce blocking in {\it time\/}
as well as space.
We simulate a process on a regular time scale measured in
terms of Monte Carlo steps (MCS). In addition to applying a standard
real-space MCRG transformation technique,
we perform a blocking of spins, by majority rule,
in consecutive discrete time steps.
In principle, the advantage of blocking in time can be two-fold. Firstly,
we expect that it will smooth out
high frequency fluctuations allowing one to reach the asymptotic
limit of the RG transformation more rapidly than with blocking in space alone.
In the same way as blocking in space iterates
away irrelevant short-length-scale behavior, we expect that blocking in
time will further eliminate short-time irrelevant memory effects.
Secondly, one can suitably adjust the
time blocking factor $b_t$ to balance the effects resulting
from blocking in space, by choosing, for example, $b_t = b^{z_o}$, where
$z_o$ is some reference exponent.  We choose $z_o=2$ for reasons
that will become clear below.

Before discussing the method in more detail, we introduce
the quantities that will be measured in time.
Critical dynamics involves
a scaling relation in which
all time scales, in the long time limit,
are measured in units of the diverging
correlation time. Thus, one needs only to choose
a convenient measure of time correlations.
We shall use the time displaced
correlation function for the magnetization $M$, as defined by
\be
\varphi_{M}(t) =  \frac{ \langle M(t_o ) M(t_o +t) \rangle -
\langle M(t_o)\rangle \langle M(t_o +t)\rangle }
{ \langle M(t_o ) - \avg{M(t_o)}^2 \rangle^{1/2}
 \langle M(t_o +t) - \avg{M(t_o +t)}^2 \rangle^{1/2}  },
\label{phi_M_eq}
 \ee
where $M$ is $\sum_i^N \sigma_i$ as usual. Other correlation
measures are discussed below. All averages
were computed from selected numerical discrete time steps $\delta t$
(usually a few MCS).
In principle, the
time-correlation function should be fitted to a series of exponentials
\be
\varphi_M(t) = \sum_i w_i e^{-t/\tau_i}
\ee
where $\tau_i > \tau_j$ if $i < j$.
However, tests made to fit the data to two exponentials
showed that $ \tau_1 \gg \tau_2$.
Indeed, all our data were well fit to a single exponential with
time constant $\tau_M$.
Data were extracted from one long simulation
from which values of $\varphi_M(t)$ were computed over a time range
of $t=0$ to $t = 5 \tau_M$.
Further averaging was also made by running 32 systems in parallel.

To calculate $z$, we use a matching procedure
\cite{Tobochnik81,Katz82,Williams85}.  In principle,
after the irrelevant variables have been iterated away, the probability
distribution function will remain invariant under further
renormalization-group transformations.
It is expected that, after a finite number
of iterations, contributions from the irrelevant variables will be
negligible.  Then, any quantity determined after $m$ blockings of an $N$
spin system should be identical to those determined after $m+1$
blockings
of a system of $Nb^d$ spins. In our new method,
time scales after $m$ blockings of the
small lattice are also explicitly rescaled by a factor $(b_t)^m$, while the
larger lattice has times explicitly rescaled by a factor $(b_t)^{m+1}$.
Unless $b_t\equiv b^z$, quantities measured
on the two lattices will {\em still} be at different times
$t$ and $t'$.  Hence, close to the fixed point, we expect a
matching condition to hold: $\varphi (N,m,t) = \varphi(Nb^d,m+1,t')$
for a correlation function $\varphi$.
{}From this, the time rescaling factor $t'/t$ can be calculated, through
the measurement of a suitable correlation function,
$\varphi_M$ in our case.  The
difference between the estimate of $z$ and the correct dynamical critical
exponent can then be obtained, since
\begin{equation}
               \frac{t'}{ t} = b^{z-z_o},
\end{equation}
where $b^{z_o}\equiv b_t$.

Our RG transformation was done in the following way.
During a simulation, every four configurations, each separated by
one MCS, were ``blocked'' in space and time:
One block spin was made by majority rule of the
16 spins coming from 4 consecutive configurations of 4
neighboring spins, ties were broken at random.
This corresponds to a space blocking
factor $b = 2$, and a time blocking factor of $b_t = 2^{z_o} = 4$.
Such a choice would give asymptotically trivial rescaling
of length and time if $z=2$, and we expect it to
make our study sensitive to the difference $(z-2)$, which is small.

Instead of doing point to point matching, i.e., matching each
discrete time step, the quality of our data is such that we have matched
the entire function $\varphi_M(t)$, since it could
be well fitted to one exponential. More explicitly,
we have $\tau_M \sim \xi^z$ and a renormalized system for which
$\tau_M' \sim \xi'^z$ where $\xi' = \xi/b$. Therefore,
without blocking in time, the critical
exponent is obtained from $\tau_M/\tau_M' = b^z$. Now, if time
is rescaled in such a way that $\tau_M'' = \tau_M'/b_t$,
then $\tau_M''/\tau_M = b^z/b_t = b^{z-z_o}$.
The discrepancy $(z-2)$ can then be obtained from
\be
z - 2 = \frac{\ln(\tau_M(L, m)) - \ln(\tau_M(bL, m+1))}{\ln b}.
\label{match}
\ee

Simulations were done on two-dimensional nearest-neighbor square lattice
spin-flip Ising systems with periodic boundary conditions.
We used a single-flip multi-system algorithm i.e, one running
different systems in parallel instead of the more
common multi-flipping one-system algorithms.
The only correlation between the parallel systems is
the sharing of the updating sequence history but we expect this to
be negligible.
The dynamic algorithm was of {\it Metropolis} type, i.e.,
one using a flipping probability based on min$(1, \exp^{-\Delta E/k_BT})$.
Systems were initialized for 20-50 $\tau_M$ at the critical
temperature of the infinite system \cite{Onsager44}. All
measurements were made at the same temperature.

The results could also be interpreted using a finite-size scaling
analysis. According to this approach, the correlation length of the
systems should be of the order of the system size so that
\be
\tau_M \sim L^z.
\ee
Therefore, the correlation time for different system sizes directly
yields the critical dynamic exponent.
Consistent results from both methods will be presented in the next
section.

\section{Results}

{}From each of the simulations detailed in table~\ref{table1}
a correlation time was extracted, as listed in table~\ref{table2}.
Figures~\ref{fi:2} and \ref{fi:3} show a typical decay
of $\varphi_M(t)$,
demonstrating that the time correlations for finite systems can be well
described by an exponential.
All the fits were done using a least-square fit algorithm.
The dynamic critical exponent could then be obtained by comparing
the values of $\tau_M$ for different systems. Values of $z$ thus
obtained are listed in table~\ref{table3}. The first striking fact
is that a point to point finite-size scaling yields a value
that compares very well with the one obtained from the dynamic
MCRG we propose. This suggests that the discrepancies in the values of $z$ as
obtained from systems of different sizes are
not systematic errors in the evaluation methods.

Although our study involves more accumulated averages
than any previous work,
it is still difficult for us to uniquely extract
errors. We have found
that the ``instantaneous'' values of $\tau_M$ contains large fluctuations.
Figure~\ref{fi:4} shows that even when binned and averaged over
$\sim 75 \tau_M$, the value of $\tau_M$ averaged for 32 systems
still contains a fair amount of fluctuations.
Therefore, as is well known \cite{Zwanzig69,Ferrenberg91},
very long simulations are required
to get a representative value of time correlations.
Furthermore, the regimes
of fitting of $\varphi_M(t)$ have been chosen as well as possible,
but it should be noted that
a logarithmic scale applied to small numbers can introduce
large fluctuations.
Nevertheless, we found that our data was well represented by
simple exponential decay
over the long times we considered ($t \le 5 \tau_M$).

If one computes the same time correlations with the absolute value
of the magnetization, then one finds much smaller correlation times,
thus showing that the ergodic time is for the most part responsible
for the large value of $\tau_M$. Since the ergodic time is due to
a finite-size effect, it is not surprising to find such a good
agreement between our results and finite-size scaling.
Also, comparison with other
kinds of time correlations \cite{Williams85}, e.g.
\[ C(t) = \sum_i \sigma_i(t_o)\sigma_i(t_o + t)\]
or
\[ E(t) = \sum_{ij} \sigma_i(t_o)\sigma_j(t_o + t), \]
where $i$ and $j$ are nearest neighbors, showed that $\varphi_M(t)$
is by far less noisy. See figure~\ref{fi:1}.

Evaluation of the data with standard finite-size scaling gives
a result of $ z = 2.14 \pm 0.01$, consistent with recent works which
used this technique \cite{Takano82,Ito87,Landau88} on systems of
comparable size.
However, pointwise scaling shows that small systems tend to show
a larger value of $z$ and that this tendency seems to disappear for
$L$ larger than $\sim 32$.
This result is somewhat
surprising since a previous MCRG study \cite{Williams85}
observed a small $z\approx 2$ from matching systems
of size $L = 8$ and $16$, and obtained a larger value when matching larger
systems.  On the other hand,
a numerical study of very large systems \cite{Mori88} gave
a  low value of $z = 2.076 \pm 0.005$. Therefore, there is still some
reason for concern on $z$'s dependence on system size.

Some authors \cite{Mazenko81} have argued that the smallness of the dynamic
critical region could be held responsible for the difficulties
encountered in evaluating $z$. However, the scaling relations
obtained near the critical point \cite{Miyashita83} do not seem to
show such a sharp region, and those effects are thought to
be strong finite-size effects. Moreover, our MCRG results
are quite consistent from the first level of RG,
as the level of iteration of RG is increased.
This self-consistently shows that the system
is in the critical regime.

Finally, by considering that small systems have a systematic error and
do not seem to be in the asymptotic scaling regime, we obtain
an estimate of $z = 2.13 \pm 0.01$ from the MCRG method we propose.
This value is consistent with most of the recent work
\cite{Takano82,Kalle84,Williams85,Tang87,Ito87,Jan88,Landau88},
as well as our finite-size scaling study herein.

\section{Conclusion}

We showed that time can be used as a renormalizable variable
in a MCRG method. Combined with standard real-space MCRG
techniques, we extracted a value for the dynamic critical
exponent that was consistent with values extracted from
the same data by finite-size scaling. Our estimated value of
$z=2.13 \pm 0.01$ is also consistent with most recent estimates.

The motivation for introducing this method was the possibility
of obtaining a more accurate value of $z$. Unfortunately, we found
that the method was not superior to conventional finite-size scaling
or real-space MCRG techniques for this problem, although
it proved to be at least
equivalent.  Further studies would be required
to see if this new MCRG method would be of use for
other related dynamic problems, such as spinodal decomposition.

\nonum
\section{Acknowledgements}
This international collaboration has been made possible by a grant from
NATO within the program \lq\lq Chaos, Order and Patterns; Aspects of
Nonlinearity", project No. CRG 890482.
This work was also supported by the Natural Sciences and Engineering
Research Council of Canada,
{\it le Fonds pour la Formation de
Chercheurs et l'Aide \`a la Recherche de la Province de Qu\'ebec\/},
and by the Supercomputer Computations Research Institute,
which is partially funded by the U.S.
Department of Energy, contract No. DE-FC05-85ER25000.


\newpage

\figure{\label{fi:1}} Different time-displaced correlation functions for
a system of
$64 \times 64$ sites. Note the large difference between $\varphi_{E}(t)$
and $\varphi_{M}(t)$ showing that the energy relaxes much more
rapidly than the order parameter. Also note the noise common
to $C(t)$ and $E(t)$. We used
$\varphi_{M}(t)$ in our estimations.

\figure{\label{fi:2}} Critical dynamics
MCRG on a $64 \times 64$ system. Curves are for $m=0,1,2,3,4$, from
top to bottom.
Averaged over 32 independent systems observed for 11 468 800 mcs.
The equilibrating time of 204 800 mcs and $\varphi_{M}$
calculated every 16 mcs.

\figure{\label{fi:3}} Critical dynamics
MCRG on a $32 \times 32$ system. Curves are for $m=0,1,2,3$, from top to
bottom.
Averaged over 32 independent systems each observed for 8 192 000 mcs.
Equilibrating time of 81 920 mcs and $\varphi_{M}$
calculated every 16 mcs.

\figure{\label{fi:4}} The evolution of the value of the cumulated average of
the time-displaced correlation functions for a system of
$32 \times 32$ sites. Even if we accumulated a large amount
of data, we see that the cumulative average still contains
a fair amount of fluctuations.

\newpage

\begin{table}
\caption{\label{table1}Simulation details for the various systems.
The range is the amount of Monte Carlo steps per spin
({\it  mcs}) for which the function $\varphi_M(t)$
has been extracted. It can be thought of as an observation window
over one simulation running in time. The total mcs for one system can be
obtained by multiplying columns 2 and 4. All the systems were first
equilibrated for 10 times the value of column 2, which in turns is of
the order of 2-5 $\tau$.}

\begin{tabular}{|c|c|c|c|c|} \hline
$L$ & range (mcs) & $\delta t$ (mcs) & average \# & $m_{\mbox{max}}$ \\ \hline
8 & 1024 & 4 &  32000 & 2 \\
12 & 2048 & 8 &  32000 & 2 \\
16 & 8192 & 16 &  32000 & 3 \\
16 & 8192 & 16 &  32000 & 3 \\
16 & 4096 & 16 &  32000 & 2 \\
20 & 8192 & 16 &  32000 & 2 \\
24 & 8192 & 16 &  26400 & 3 \\
24 & 16384 & 16 &  32000 & 3 \\
32 & 16384 & 16 &  32000 & 4 \\
32 & 16384 & 16 &  32000 & 4 \\
48 & 32768 & 16 &  4832 & 4 \\
64 & 65536 & 64 &  18016 & 4 \\
96 & 98304 & 64 &  2336  & 4 \\ \hline
\end{tabular}
\end{table}

\bigskip

\begin{table}
\caption{\label{table2}The values of $\tau_M$ as estimated from
a least square fit of the time-time correlation function $\varphi_M(t)$
to a simple exponential.
The values are in {\it mcs} and the simulation characteristics
can be read from the respective entry in table~1.
The errors indicated in parenthesis are those obtained
from the fit. The last line represents the value of $z$ obtained
from finite-size scaling analysis applied to each column.}

\begin{tabular}{|c|c|c|c|c|c|} \hline
$L$ & \multicolumn{5}{c|}{MCRG iteration number} \\ \hline
   &   0     &    1    &    2    &   3   &   4  \\ \hline
8  &  261.41(6) &   64.31(4) &   16.34(4) &          &      \\
12 &   631.8(2) &  157.9(1)  &   39.5(1)  &          &      \\
16 &  1193.9(5) &  298.4(2)  &   74.45(4) & 18.6(1)  &      \\
16 &  1179.3(3) &  294.6(2)  &   73.5(1)  & 18.5(1)  &      \\
16 &  1189.1(5) &  296.9(3)  &   74.3(2)  &          &      \\
20 &  1930.7(3) &  482.9(1)  &  120.5(2)  &          &      \\
24 &  2853.9(3) &  716.0(4)  &  179.0(3)  &  44.8(2) &      \\
24 &  2837.9(3) &  713.3(6)  &  178.3(3)  &  44.5(1) &      \\
32 &  5366(2)   &  1341.9(6) &  335.4(3)  &  84.0(2) & 20.7(2) \\
32 &  5343.4(5) &  1335.6(3) &  333.8(2)  &  83.4(1) & 20.8(1) \\
48 & 12528(2)   &  3130.4(5) &  782.8(3)  & 195.6(1) & 49.4(2) \\
64 & 23316(4)   &  5814(2)   & 1453(1)    & 363.3(8) & 92(1)   \\
96 & 54620(20)  & 13660(10)  & 3415(5)    & 852(3)   & 214(2)  \\ \hline
$z$  & 2.153     &  2.145    &  2.142    &  2.125  & 2.128  \\ \hline
\end{tabular}
\end{table}

\bigskip

\begin{table}
\caption{\label{table3}Values of $z$ as estimated from
point to point finite-size scaling (FSS) and
by matching the correlation times according to
relation~\refp{match}.}

%

\bigskip

Figure 4\\
\clearpage

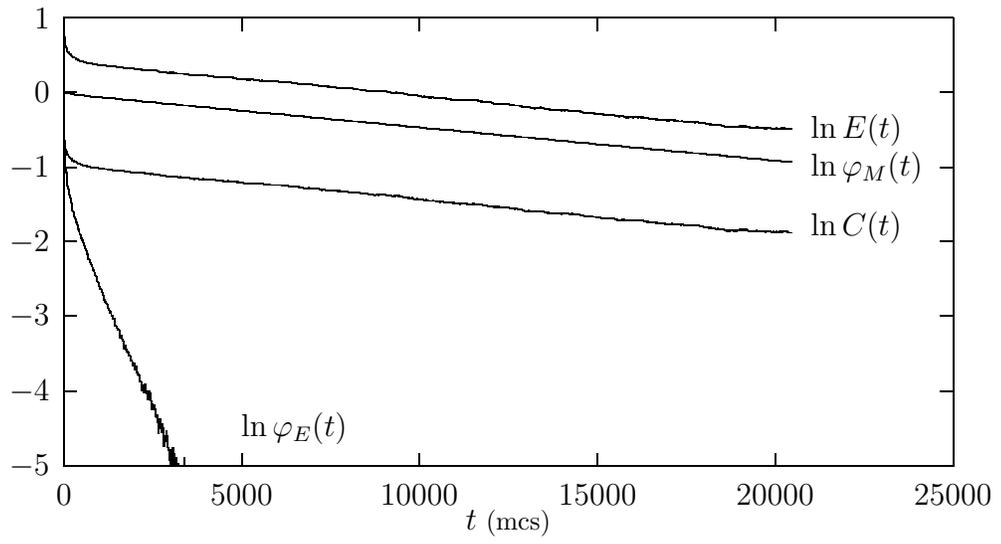
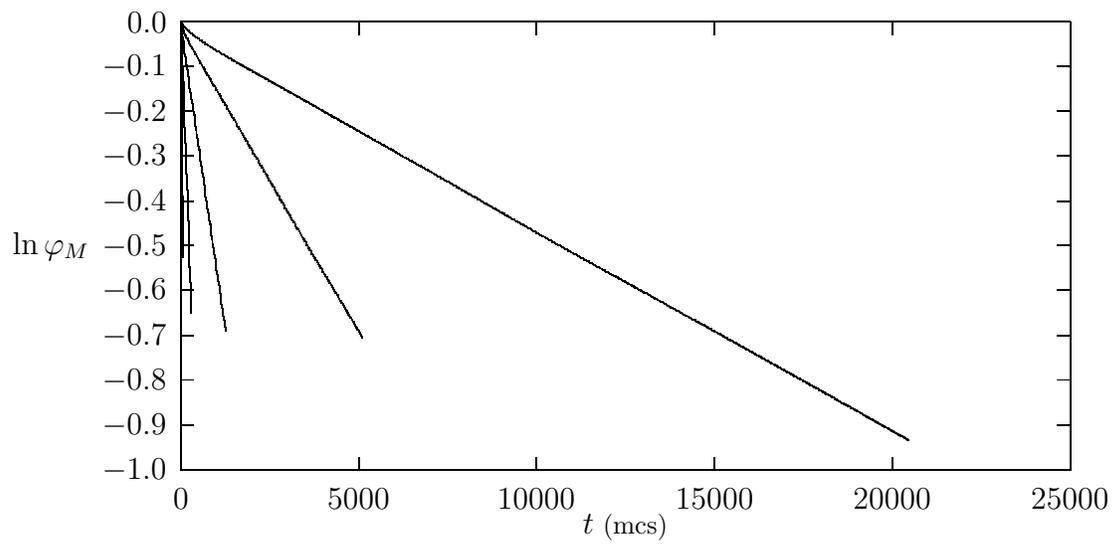
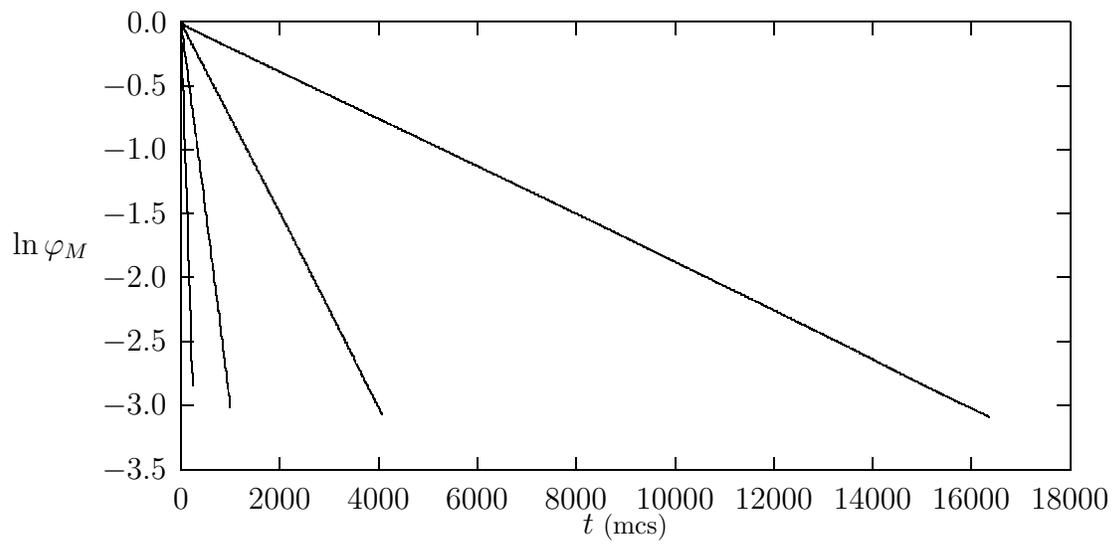
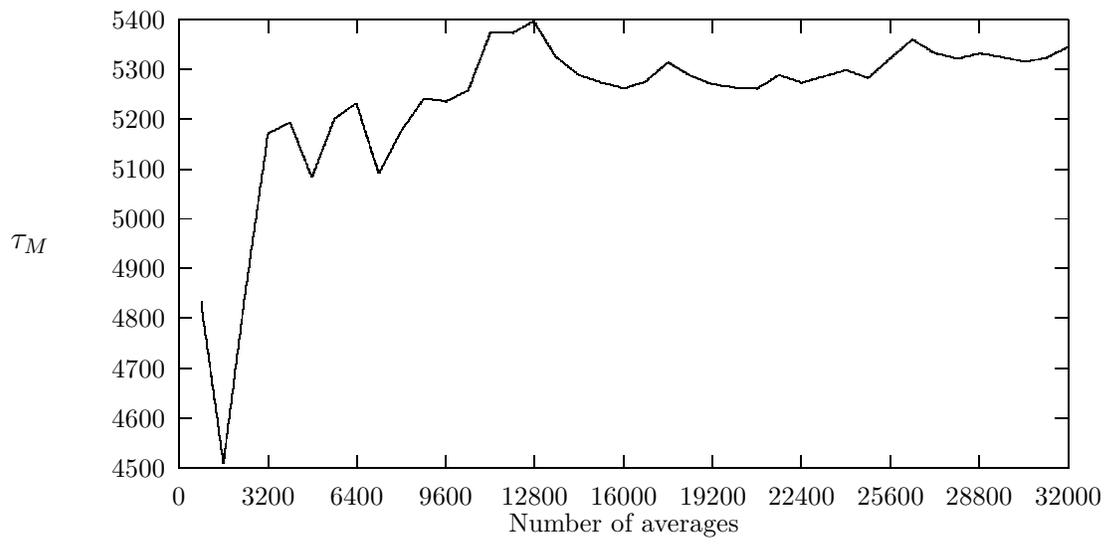
\end{document}